%% file: Simon-arXiv-21.tex
\documentclass[11pt]{article}

\usepackage{graphicx}
\graphicspath{{./Data/}}
\DeclareGraphicsExtensions{.pdf,.png,.jpg}

\usepackage[margin=1in]{geometry}

\usepackage{setspace}
\begin{document}

\title{Drop Interface and Airflow Unsteadiness in\\ Wind-Forced Drop Depinning}
\author{Roger L. Simon, Jr. \quad Edward B. White\thanks{ebw@tamu.edu}\\[1ex]
Department of Aerospace Engineering\\
Texas A\&M University}
\date{\relax}
\maketitle

\input{abstract}
\input{intro}

\input{approach}

\input{results}
\input{conclusions}

\section*{Acknowledgements}

The authors wish to thank Sungyon Lee and Alireza Hooshanginejad
for multiple fruitful conversations regarding this work.
The authors also with to thank the U.S. National Science Foundation
for support from grants CBET-1839103 and HRD-1810995.

\bibliographystyle{unsrt}
\bibliography{Simon}

\end{document}

%% file: abstract.tex
\section*{Abstract}

Liquid drops that are pinned to solid surfaces by contact-angle
hysteresis can be dislodged by wind forcing. When this occurs
at high Reynolds numbers, substantial drop-interface oscillations
precede depinning. It has been hypothesized that coupling between
drop interface oscillations and unsteady airflow vortices are
important to the depinning process. This possibility is investigated
using simultaneous high-speed side-view drop images and airflow
fluctuation measurements.  The results show no evidence of coupling
across a range of drop volumes and wind speeds for water drops in
air. When properly scaled by drop volume, drop interface fluctuation
frequencies are not affected by wind speed. Airflow vortex shedding
occurs as if the drop were a solid surface protuberance. For the
air/water system in these experiments, vortex shedding frequencies
are substantially higher than drop interface frequencies and the disparate
frequencies may make the hypothesized coupling impossible.

%% file: intro.tex
\section{Introduction}

Liquid drops that rest on solid surfaces are pinned in place by
contact-angle hysteresis unless external forcing exceeds the maximum
available pinning force. Beyond the maximum force, drops depin and run back along the surface.
Early experimental work on depinning by gravity on inclined surfaces
was performed by Macdougall and Ockrent \cite{Macdougall-PRSLA-42},
Bikerman \cite{Bikerman-JCS-50}, and Furmidge \cite{Furmidge-JSC-62}.
These studies established the maximum force a drop can resist scales as
$\gamma w (\cos\theta_r -\cos\theta_a)$ where $\gamma$ is surface
tension, $w$ is the drop width, and $\theta_r$ and $\theta_a$ are
the receding and advancing contact angles, respectively. This result
is straightforward to establish because gravity forcing is simply
the drop weight times the sine of the surface inclination. Moreover,
the gravity-forced drop shape is steady and $\theta_r$ and $\theta_a$ can be
observed using side-view cameras.

It is of interest to understand drop depinning by wind forcing, a
much-more complex phenomenon. When the airflow velocity is low or
the drops are very small relative to the airflow boundary layer,
drops may depin due to viscous shear. An analytical study in this
regime based on lubrication theory was developed by
Dussan~V.~\cite{Dussan-JFM-87}. When drops are larger, flow over
the drop separates and forcing is mainly due to the pressure
difference between the windward and leeward side of a drop
\cite{Durbin-JFM-88,Hooshanginejad-PRF-17}. Lee-side separation
occurs when the Reynolds number based on the drop height, $Re_h$,
is large. This Reynolds number is defined $Re_h = \rho_{\mathrm{air}}\,
U h/\mu_{\mathrm{air}}$ where $\rho_{\mathrm{air}}$ and
$\mu_{\mathrm{air}}$ are air density and viscosity, respectively.
In this situation, depinning occurs when a critical Weber number,
$W\!e=\rho_{\mathrm{air}}\, U^2 h/\gamma$, is exceeded
\cite{Durbin-JFM-88,Milne-Langmuire-09,White-PRF-21}.  In the
definitions of $Re_h$ and $W\!e$, $U$ is taken to be the wind velocity
at the maximum drop height, $h$, in the undisturbed air flow.

Studies on wind-forced drop depinning are extensively reviewed by
Milne and Amirfazli \cite{Milne-Langmuire-09}. Wind-forced drops are
difficult to model because the drag force applied by
wind depends on the shape of the drop, drop size relative to the
airflow boundary layer, and the laminar or turbulent state of the
flow. There can also be substantial unsteadiness of both the airflow
and drop interface shape. Milne et al.~\cite{Milne-ACIS-14} reviewed
a variety of studies on drop interface oscillations and performed
wind-forced drop-oscillation studies to evaluate the success of
drop oscillation models. A two-dimensional simulation by Lin and
Peng \cite{Lin-HTAR-09} suggests a coupling between drop shape
oscillations and airflow vortex shedding by the drop is the cause
of drop shape unsteadiness.

Wind-forced drop experiments by Milne et al.~\cite{Milne-ACIS-14} and
Esposito et al.~\cite{Esposito-JPS-10}, both observed that drop
oscillation frequencies do not depend on wind velocity but decrease
with increasing drop volume as $V^{-1/2}$. Milne et al.\ also showed
that mode shapes and natural frequencies of wind-forced drops are
similar to those excited by other forcing modalities and specifically
studied oscillations corresponding to an analysis by Chiba et
al.~\cite{Chiba-JSV-12}. Those authors analyzed hemispherical drops at
zero Bond number with pinned contact lines and predicted nondimensional
natural frequencies of $f^* = f\,(\rho_{\mathrm{water}}\,V/\gamma)^{1/2}
= 0.51$ for mode (1,1), 1.02 for mode (0,1), 1.12 for mode (2,1),
and 1.72 for mode (1,2).  (Chiba et al.\ give different numerical
values than these because their nondimensional scheme is based on
circular frequency and drop radius cubed.) The lowest-frequency
mode (1,1) is a longitudinal downwind/upwind oscillation. Mode (0,1)
is an axisymmetric vertical oscillation and mode is (2,1) a shape
that alternates between elongation in the longitudinal and lateral
directions. In the $(m,n)$ mode nomenclature, $m$ is the meridional
mode number ($m=0$ is axisymmetric) and $n$ is the azimuthal mode
number. Milne et al.~\cite{Milne-ACIS-14} explain that modes (0,1)
and (2,1) would be difficult to distinguish using side-view images
because their profile shape oscillations would appear similar and
their frequencies are similar.

No studies have yet considered how drop unsteadiness may change
near critical depinning conditions or drive the depinning event.
Substantial drop-interface oscillations have been observed to precede
critical conditions \cite{White-PRF-21,White-JFE-08,Seiler-PRF-19}
but the character of the oscillations was not the object of those
studies. The work by White and Schmucker \cite{White-PRF-21} in a
laminar boundary layer found that depinning always occured at $Re_h$
values known to produce unsteady airflow vortex shedding
\cite{Acarlar-JFM-86}. Combined with the simulation results by Lin
and Peng \cite{Lin-HTAR-09}, this suggests a coupling between unsteady
drop interface motion and vortex shedding may be a key feature of
the depinning process.

Vortex shedding by solid surface protuberances has been extensively
studied in the  aerodynamics literature. A seminal experimental
study of hemispheres and teardrop shapes in boundary layers is by
Acarlar and Smith \cite{Acarlar-JFM-86}. Teardrop shapes resemble
wind-forced drops except have a different shape for the
``receding'' part of their perimeter than do wind-forced drops. (Compare Figure
7a of Ref.~\cite{White-PRF-21} to Figure 2b of Ref.~\cite{Acarlar-JFM-86}.)
Acarlar and Smith found teardrop-shaped protuberances with $400 <
Re_h < 1400$ shed periodic airflow vortices. The shedding frequency,
$f$, is described by a Strouhal number $St = f h/U$ that increases
from about 0.15 to 0.3 across that range of $Re_h$. Shedding is
less regular at higher $Re_h$ values.

Whether vortex shedding frequencies may be altered when the
protuberance is a liquid drop rather than a solid is not known. It
could be that a coupling between drop oscillations and vortex
shedding leads to combined aerodynamic and inertial forces that
overcome the maximum surface-tension pinning force. What's more,
because those forces are highly unsteady, this may result in drop
``shuffling'' motion observed by Saal et al.~\cite{Saal-PRF-20} and
others. With this in mind, the particular goal of this study is to
make simultaneous measurements of drop interface motion and airflow
unsteadiness at wind speeds approaching drop depinning conditions.
Side-view drop images are recorded using a high-speed video camera.
Airflow unsteadiness is measured using a hotwire anemometer located
in the drop wake. The resulting data is intended to reveal whether
a coupling phenomenon occurs between the wind vortices and drop
oscillations that may trigger depinning.

%% file: approach.tex
\section{Approach}

Measurements were made of water drops on a roughened aluminum
substrate in a small wind tunnel designed for studies of wind-driven
drop depinning \cite{White-PRF-21,Schmucker-PhD-12}. The wind-tunnel
test section is 250~mm long by 25~mm tall by 50~mm wide. The sides
and top wall are transparent acrylic. Interchangeable aluminum
surface samples are located flush with the test section floor. The
design allows the flow to cross onto the substrate smoothly and
allows top- and side-view cameras to image the surface without
obstruction.

Laminar airflow in the test section is provided by several flow-quality treatments.
The flow enters the tunnel through a 200-mm-wide inlet fairing;
passes through a honeycomb flow straightener, two stainless-steel
turbulence screens, and a 250-mm-long, 4:1 contraction before
entering the test section. Downstream of the test section, a $5.5^{\circ}$
half-angle diffuser leads to a 80-mm-diameter fan. The small angle
prevents diffuser separation that could lead to low-frequency
unsteadiness. Wind-speed is automatically controlled using a
feedback-controlled motor that maintains the desired test velocity.

The boundary-layer displacement thickness where drops are located varies as $\delta^*
= a\,U^{-1/2}$ where $a = 1.7\ \mathrm{mm(m/s)}^{1/2}.$ Test speeds
varied from 5.9 to 9.9 m/s so $\delta^*$ values varied from about
0.5~to 0.7~mm. Initial drop heights without wind forcing ranged from
2.6 to 3.2~mm. Thus, drops can be considered to exist in nearly uniform
flow. Velocity fluctuations are approximately 0.5\% of the test-section
centerline velocity.

\begin{figure}\center
\includegraphics[width=5in]{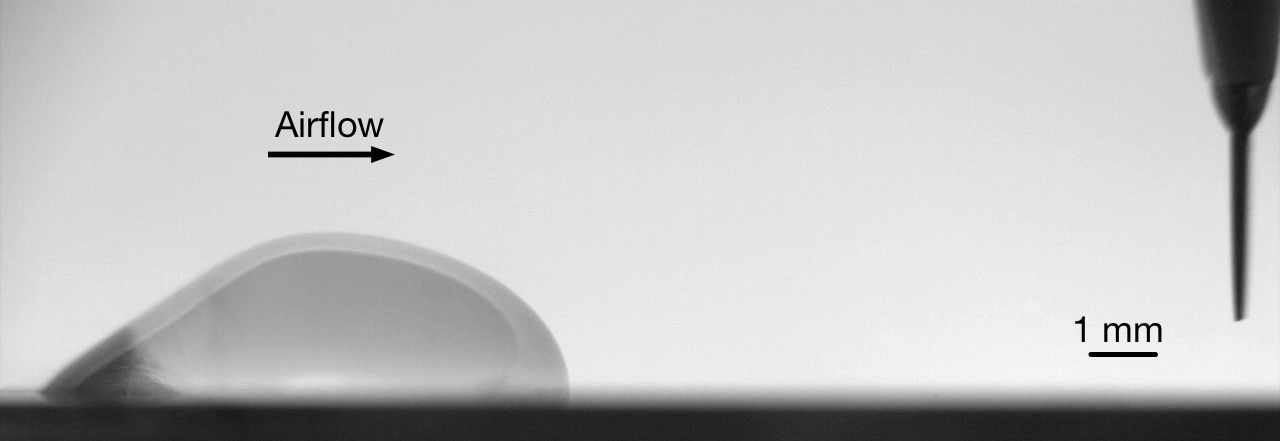}
\caption{Example high-speed video image of a 175~$\mu$L drop exposed to
7.9~m/s wind flow.}
\label{==fig:example-image==}
\end{figure}

Side-view drop images are recorded using a
Chronos 1.4 high-speed camera that recorded 2~seconds of $1280\times1024$
video per test point at 500 frames/sec. The camera was mounted with
its optical axis aligned with the level of the aluminum
substrate. Figure~\ref{==fig:example-image==} shows an example image
obtained using this system. The hotwire anemometer used to measure
airflow unsteadiness in the drop wake is visible downwind of the
drop.

\begin{figure}\center
\includegraphics[width=3in]{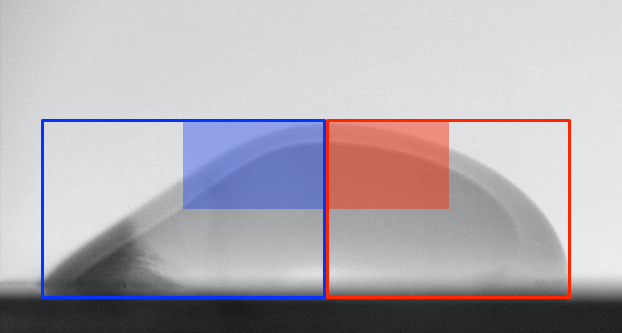}
\caption{Video analysis regions of interest for the drop
pictured in Figure~\ref{==fig:example-image==}. The filled red rectangle
is the advancing region; the filled blue rectangle is the receding region.}
\label{==fig:roi==}
\end{figure}

Milne et al.~\cite{Milne-ACIS-14} developed a sophisticated approach
for analyzing drop interface motions on a mode-by-mode basis. A
simplified alternative method is used here. Of particular interest
are the relative amplitudes of the three lowest-frequency modes
(0,1), (1,1), and (2,1). The work by Chiba et al.~\cite{Chiba-JSV-12}
suggests that mode (1,1) is the lowest frequency and, from a side
view, would manifest as a left/right motion of the drop interface.
Modes (0,1) and (2,1) would occur at about twice the frequency of
mode (1,1) and would appear as an up/down motion of the interface from a
side view. Two separate regions of interest are evaluated in order
to distinguish these motions. These are indicated by the red and
blue shaded regions in Figure~\ref{==fig:roi==}. Each region is defined
as a rectangle that begins at the drop apex and extends to one half
the vertical and horizontal distance to the advancing (red) or
receding (blue) contact point in the side-view image. As the
experiment proceeds, a new apex point is selected for each wind
speed as the drop deforms under increasing forcing.

The mean grayscale pixel intensity in both regions of interest are
calculated for each video frame. Because the drop appears darker
than the background, both regions' mean value would decrease (more
dark) with upward vertical motion; the advancing region would
decrease as the interface moves right; and the
receding region would decrease as the interface moves
left. With this arrangement, the sum of the red- and
blue-region time signals would tend to indicate vertical motion
while the difference would tend to indicate lateral motion. The
signal amplitudes are simply proxies for interface motion and do
not correspond to a physical mode amplitude. Nevertheless, the frequencies
and relative signal amplitudes are intended to reveal how interface
motion changes as wind speed increases.

Instantaneous wind velocities are measured simultaneously with video
recordings using a constant-temperature hotwire anemometer sampled
at 3000 Hz. The hotwire sensor is 1 mm long, is 5 $\mu$m in diameter, and
is oriented perpendicular to the plane of Figure~\ref{==fig:example-image==}. It is
supported at the bottom of the thin, vertically oriented stainless-steel
prongs visible in the figure.

The aluminum test surface was cleaned with acetone before each drop was
applied. After it dried, a drop was carefully applied using a graduated
syringe with 1~$\mu$L resolution. At the start of the experiments the drops had
near-circular contact lines. The experiments proceeded automatically
with the control program slowly increasing wind speed until
pre-determined velocities were achieved in the test section. Once
a steady, on-condition wind speed was reached, hotwire and video
recording was automatically triggered. Each drop experiment included multiple
velocity set points depending on the anticipated depinning wind
speed.

%% file: results.tex
\section{Results}

Drop volumes from 100 to 250~$\mu\mathrm{L}$ were
studied at wind speeds of 5.9, 7.9, and 9.9~m/s. Larger drops
depinned at less than the maximum test wind speed: 125~$\mu\mathrm{L}$
and higher at less than 9.9 m/s and 200 $\mu\mathrm{L}$ and higher
at less than 7.9 m/s. Without wind forcing these drops had maximum
heights ranging from $h=2.6$ to 3.2~mm, corresponding to Bond numbers
of $Bo = \rho_{\mathrm{water}}\,g h^2/\gamma = 0.9$ to 1.4. The Bond
number varies less than $V^{2/3}$ because when $Bo$ exceeds about
1.0, adding additional volume tends to increase the contact line
diameter rather than drop height. (See Ref.~\cite{White-PRF-21} for
additional details for this particular system.) $Re_h$ values vary from
1020 to 1710 for $h=2.6$ mm across the three speeds and from 1020 to
1240 for 5.9~m/s wind across the volume range.

\begin{figure}\center
\includegraphics[width=4.5in]{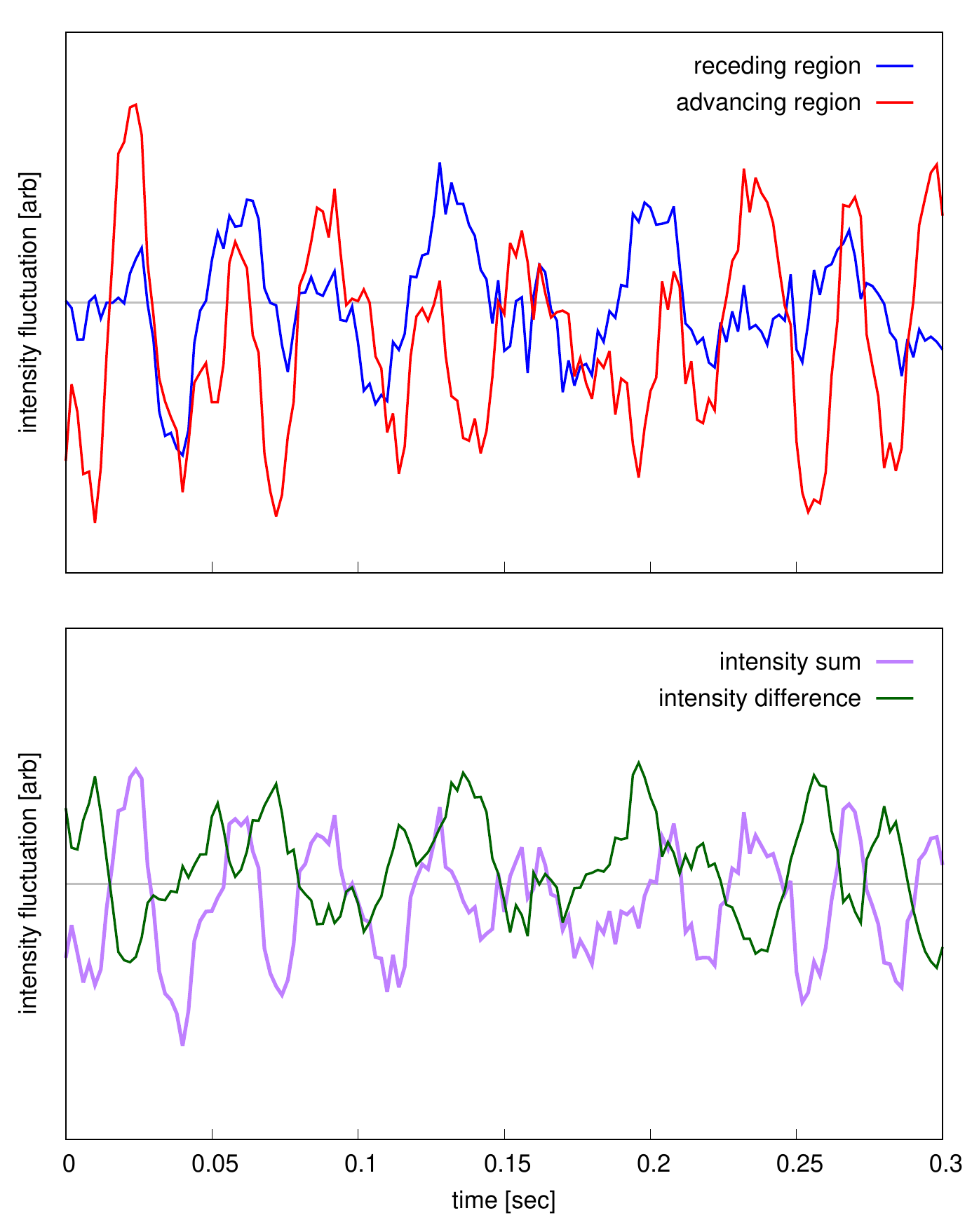}
\caption{Example pixel-intensity time series of a
125~$\mu\mathrm{L}$ drop forced by 5.9~m/s wind.}
\label{==fig:intensity-time-series==}
\end{figure}

\begin{figure}\center
\includegraphics[width=4.5in]{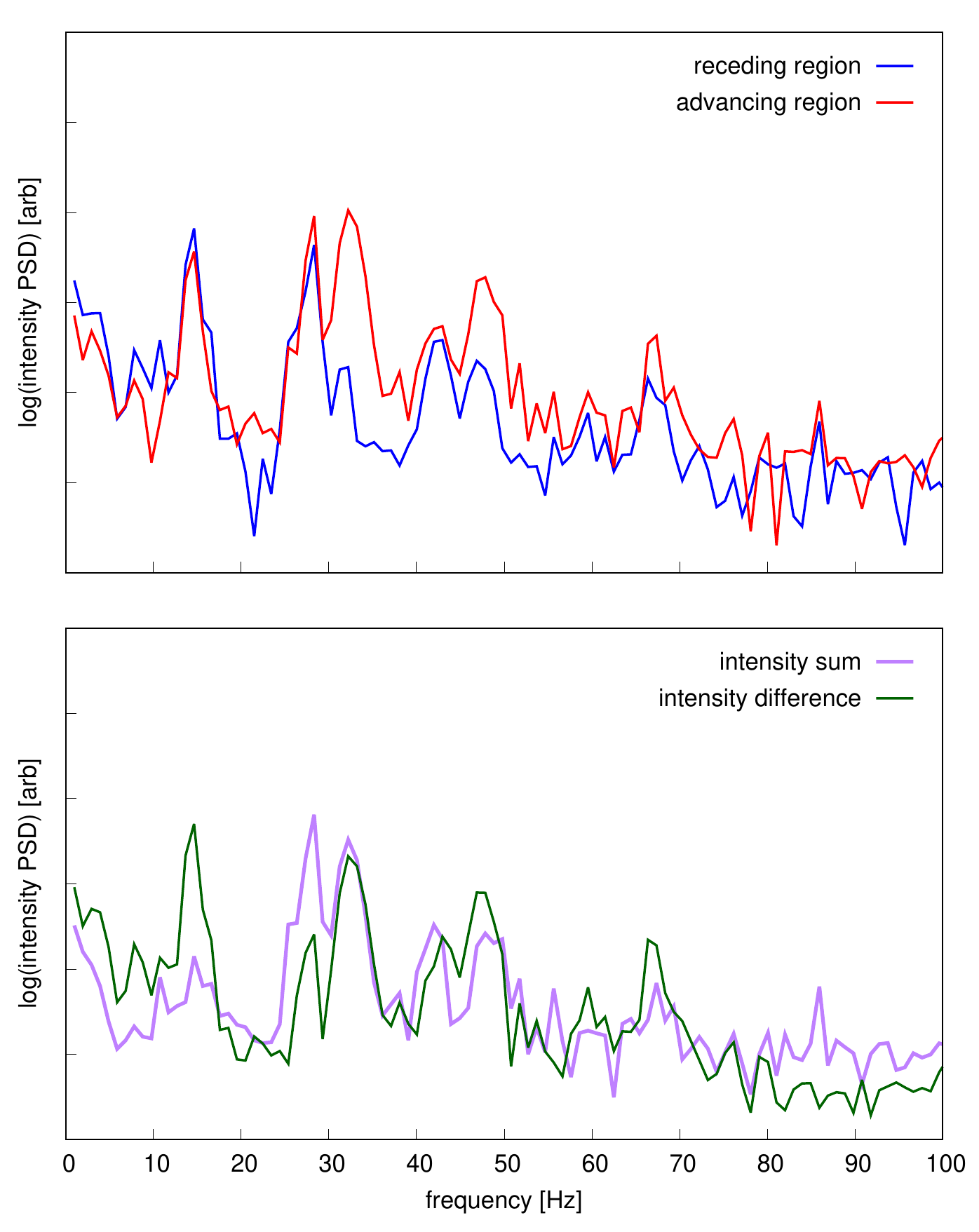}
\caption{Power spectral densities of signals of 
Figure~\ref{==fig:intensity-time-series==}.}
\label{==fig:psd-1==}
\end{figure}

Example time signals of mean pixel intensities of a $125\ \mu\mathrm{L}$ drop
forced by 5.9 m/s wind are shown in Figure~\ref{==fig:intensity-time-series==}.
The top plot corresponds to signals in the advancing and receding regions
(see Figure~\ref{==fig:roi==}) and the bottom plot includes the
corresponding sum and difference signals. Numerical values are not
given on the vertical axis because these are arbitrary pixel
intensities that do not correspond to a physically meaningful value.
Power-spectral densities (PSDs) of the same signals are shown in
Figure~\ref{==fig:psd-1==}.

Figure~\ref{==fig:psd-1==} shows prominent peaks at 14.7, 28.3, 32.2,
42.0, 47.9, and 66.4~Hz that correspond to $f^*=0.61$, 1.18, 1.34,
1.74, 1.99, and 2.76. The lowest of these frequencies correspond
reasonably closely to the lowest $f^*$ values predicted by Chiba
et al.~\cite{Chiba-JSV-12} for hemispherical drops.
Notably, the lowest frequency is significantly more
pronounced in the subtracted signal as compared to the summed signal.
This suggests the red and blue regions are fluctuating out of phase,
exactly what would be expected for the lateral oscillations of mode
(1,1) and this is the lowest-frequency mode predicted by Chiba et
al. The next two higher frequency peaks appear as a double peak
that is somewhat more pronounced in the summed signal. These appear
to correspond to the closely spaced mode-(0,1) and mode-(2,1)
fluctuations predicted by Chiba et al.

\begin{figure}\center
\includegraphics[width=4.5in]{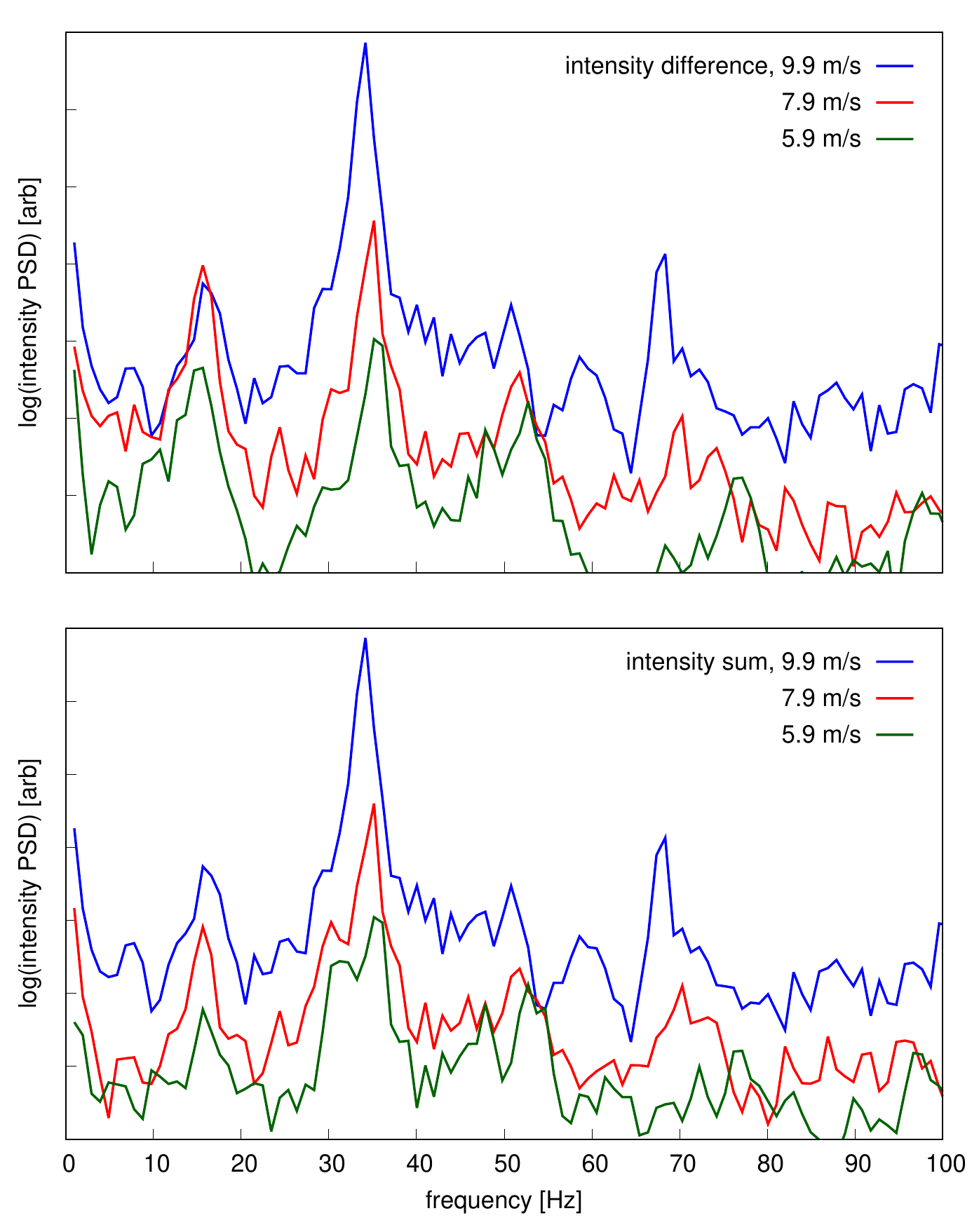}
\caption{Power spectral densities of signals of a
100~$\mu\mathrm{L}$ drop at different wind speeds.}
\label{==fig:psd-2==}
\end{figure}

As wind speed increases the prominent frequencies of a particular
drop do not change. Figure~\ref{==fig:psd-2==} shows the sum and
difference PSDs for a $100\ \mu\mathrm{L}$ drop at 5.9, 7.9, and
9.9~m/s wind speeds. Each peak occurs at a slightly higher frequency
than the corresponding peaks in Figure~\ref{==fig:psd-1==} because 
the drop volume is less. The key peaks in the difference signal are
at 15.6, 35.2, 47.9, and 52.7~Hz. An additional peak is observed
in the sum signal at 31.7~Hz, especially at the lowest wind
speed. The corresponding $f^*$ values are consistent with the $125\
\mu\mathrm{L}$ drop values: $f^*=0.58$, 1.18, 1.31, 1.78, and 1.96.
The fluctuation amplitudes increase substantially as wind speed is
increased. This is consistent with previous observations of increasing
interface unsteadiness as wind-forcing increases
\cite{White-PRF-21,White-JFE-08,Seiler-PRF-19}.  Oscillations
at 35.2 Hz increase somewhat more than other frequencies as critical
conditions are approached. However, the relative amplitudes
of the different spectrum peaks remain mostly unchanged with
increasing wind speed. So, at least within the limitations of the
current analysis technique, it does not appear as if a specific
mode emerges to drive depinning.

\begin{figure}\center
\includegraphics[width=4.5in]{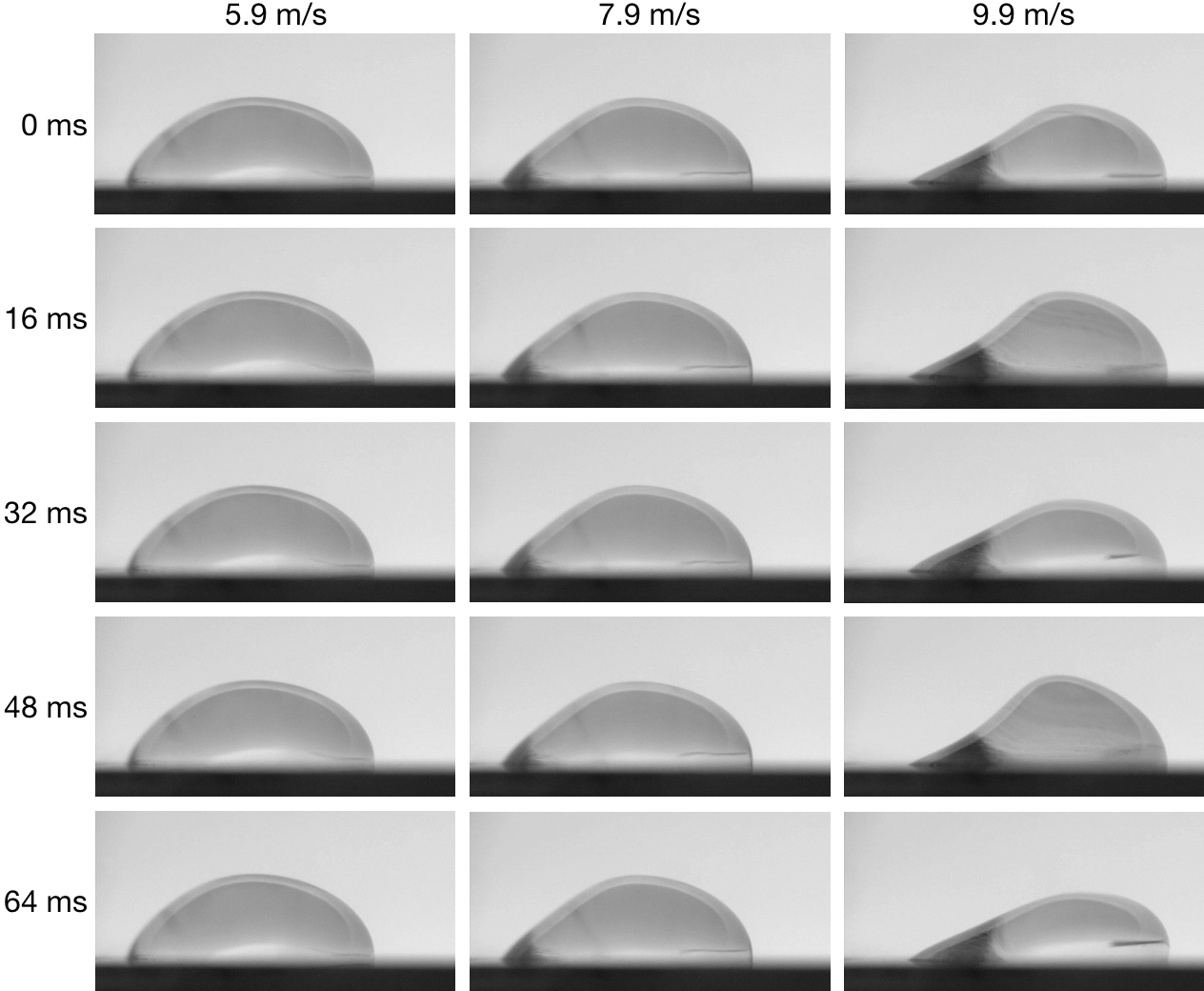}
\caption{Side-view drop images of a 100~$\mu\mathrm{L}$ drop at different wind speeds. The unforced drop height is 2.6~mm. Wind forcing is left to right.}
\label{==fig:composite-results==}
\end{figure}

Side-view images of the 100 $\mu L$ drop that produced the data in
Figure~\ref{==fig:psd-2==} are given in
Figure~\ref{==fig:composite-results==}. Progressing from 5.9~to 9.9~m/s
shows the increasing difference between advancing and receding contact
angles as wind forcing increases. Almost no interface motion can be
observed at lower speeds but oscillations are obvious at 9.9~m/s. The
lowest-frequency oscillation is 15.6~Hz so undergoes one period in 64~ms.
The next two frequencies are 31.7~Hz and 35.2~Hz that correspond to 32~ms
and 28~ms periods, respectively. That is, the figure captures one period
of the lowest frequency and approximately two periods of the next two
frequencies. The vertical interface motion associated with the
large-amplitude 35.2~Hz oscillation is depicted very clearly at 9.9~m/s in
which the large upward displacement occurs at $t=16$ and 32~ms and lower
heights occur at the opposite phase, $t=0$, 32, and 64~ms. The left/right
motion associated with the 15.6~Hz oscillation is not as clear but at
9.9~m/s can be observed by comparing the 32~and 64~ms images. The 64~ms
image shows a shift to the right relative to the 32~ms image at
essentially the same phase in the higher-frequency vertical motion.

\begin{figure}\center
\includegraphics[width=4.5in]{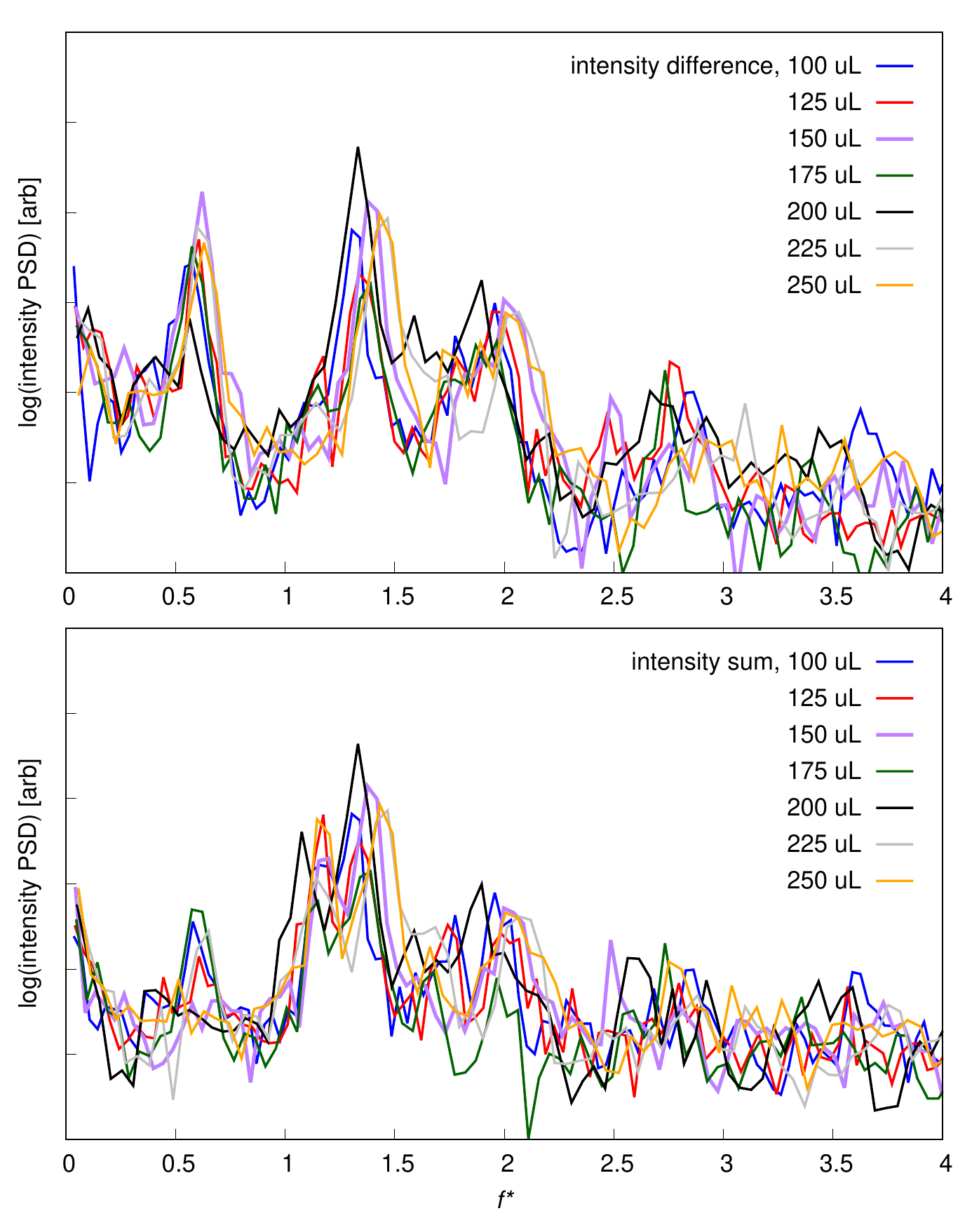}
\caption{Power spectral densities of various drop
volumes forced by 5.9~m/s wind.}
\label{==fig:psd-3==}
\end{figure}

\begin{figure}\center
\includegraphics[width=4.5in]{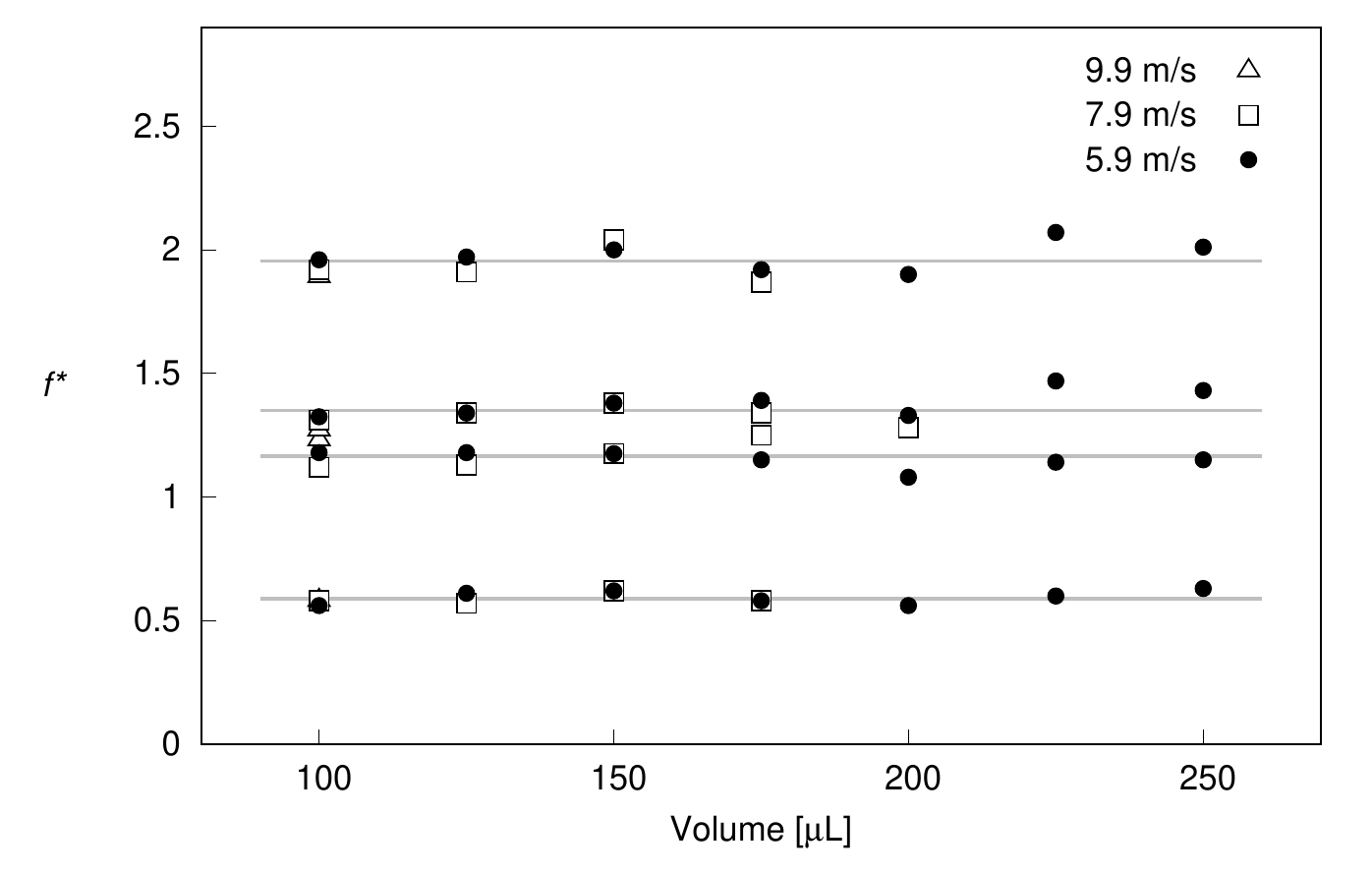}
\caption{Nondimensional oscillation
frequencies $f^*$ remain constant across drop volume and wind speed.}
\label{==fig:freq-v-volume==}
\end{figure}

The success of the $f^* = f(\rho_{\mathrm{water}} V/\gamma)^{1/2}$
scaling in collapsing the drop oscillation response is evident in
Figure~\ref{==fig:psd-3==} that shows the responses of drops from $100\
\mu\mathrm{L}$ to $250\ \mu\mathrm{L}$ at a wind speed of 5.9~m/s.
The sum and difference intensity spectra are remarkably similar
across these tests. Frequencies of the various peaks are plotted
as a function of drop volume and wind speed in Figure~\ref{==fig:freq-v-volume==}.
The lines in that figure indicate the mean $f^*$ values of each of the
spectral peaks, $f^*=0.59$, 1.17, 1.35, and 1.95. Not all peaks are
clearly visible at all speeds and volumes and only peaks that could be
clearly identified are shown. Across the different volumes and speeds,
the $f^*$ values are quite consistent. Considering the difference in
shape between the 100~and 250~$\mu\mathrm{L}$ drops and the difference
in shape as the critical depinning wind speed is approached, this is a
rather surprising result. The $f^*$ values measured here are somewhat
higher than corresponding values predicted by Chiba et
al.~\cite{Chiba-JSV-12}: 0.51, 1.02, 1.12, and 1.72. The difference is likely
because of the different initial drop shape in this experiment as
compared to the hemispherical shape and zero Bond number analyzed by
Chiba et~al.

\begin{figure}\center
\includegraphics[width=4.5in]{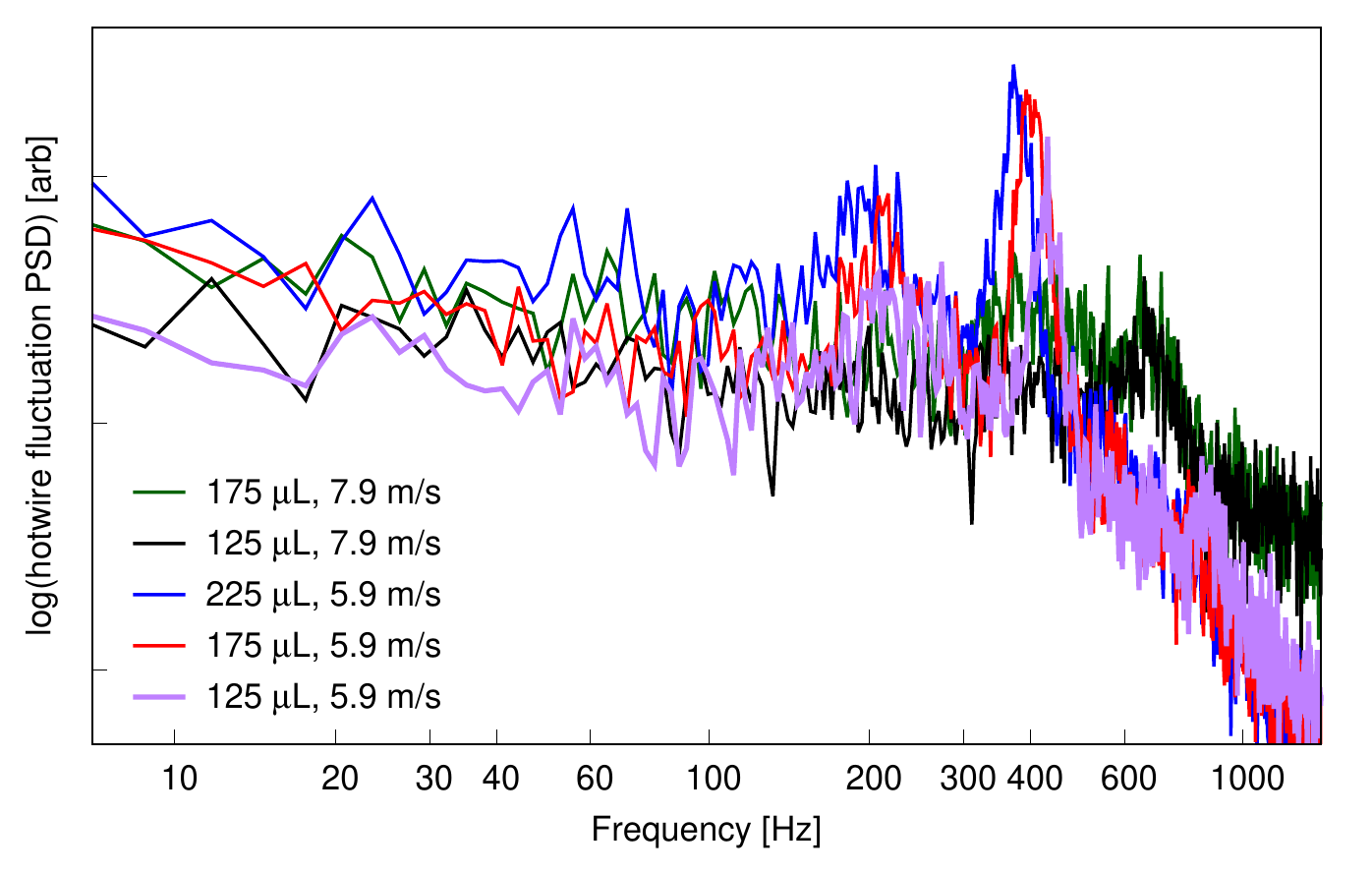}
\caption{Power spectral densities of airflow
fluctuation intensities behind various drops at different wind speeds.}
\label{==fig:psd-4==}
\end{figure}

The key question of this work is whether coupling between airflow
velocity fluctuations and drop interface fluctuations may drive drop
depinning. Hotwire measurements of airflow fluctuations downstream
of the drop show no evidence of such a mechanism. Figure~\ref{==fig:psd-4==}
shows the power spectral density of wind velocity fluctuations measured
by the hotwire about 15~mm downstream of the drop center and 1.2~mm
above the surface. None of the frequencies associated with drop interface
oscillations are observed to have elevated signal
power that would indicate the airflow is affected by drop motion.
Instead, the key features of Figure~\ref{==fig:psd-4==} are the large-amplitude
peaks at 370, 400, and 430~Hz for the 225, 175, and 125~$\mu\mathrm{L}$
drops, respectively, for the 5.9~m/s wind speed. These correspond
to $St=0.20$ vortex shedding, consistent with results by Acarlar and Smith
\cite{Acarlar-JFM-86} for solid protruberances with equivalent $Re_h$
as the drops, 1100 to 1240. At 7.9~m/s the peaks are broader and
range from about 600 to 800 Hz for both the 175 and 125~$\mu\mathrm{L}$
drops. These correspond to $St$ values from 0.23 to 0.29.  The
broader peaks at a higher Strouhal number are expected at the higher
Reynolds numbers of these drops, 1470 and 1540. Varying the hotwire
position would not change the measured frequencies but could influence
signal magnitudes. However, because the fluctuation amplitudes are
arbitrary, measurements at other downstream positions would not yield
meaningfully different results.

%% file: conclusions.tex
\section{Conclusions}

This study seeks to determine whether coupling between unsteady
aerodynamic forcing and unsteady drop interface shapes may contribute to
drop depinning. Previous studies \cite{White-PRF-21,White-JFE-08,Seiler-PRF-19}
have observed large-amplitude
interface unsteadiness prior to depinning and depinning occurs at
Reynolds numbers at which unsteady vortex shedding occurs
\cite{White-PRF-21,Acarlar-JFM-86}. Thus, a coupling between airflow
vortex shedding from the drop and drop-interface oscillations might
occur and be a central aspect of drop depinning. A coupling of this type
was reported by Lin and Peng \cite{Lin-HTAR-09} based on a two-dimensional
simulation and that finding partly motivates the current study.

To investigate whether a coupling exists, side-view video images of
wind-forced drops were made simultaneously with hotwire anemometer
measurements of airflow fluctuations in drop wakes. The video images
were analyzed using a simple spatial average of pixel intensities in
regions on the advancing and receding sides of the drop. The two regions
were selected to provide a qualitative assessment of in-phase and
out-of-phase interface motion on the two sides of the drop.

The results show strong responses at frequencies close to those
predicted for modes (1,1), (0,1), (2,1), and (1,2) by Chiba
et al.~\cite{Chiba-JSV-12} for hemispherical drops at zero Bond number.
The nondimensional frequency $f^* = f(\rho_{\mathrm{water}}\,
V/\gamma)^{1/2}$ at which these oscillations appear was unaffected by
drop volume and wind speed across the range of values considered in the
present experiment. Milne et al.~\cite{Milne-ACIS-14} and Esposito et
al.~\cite{Esposito-JPS-10} observed the same behavior.

Similar to previous observations, drop interface motion was observed to
increase dramatically as critical wind speed was approached. The general
shape of power spectra representing interface motion did not change with wind speed.
Hotwire measurements show no evidence of wind-velocity fluctuations at key drop oscillation
frequencies. Instead, the fluctuations are exactly what is expected for
solid protuberances \cite{Acarlar-JFM-86}. It may be the case that vortex
shedding frequencies are so much higher than drop interface frequencies that
no possibility of coupling exists. That is, on the timescale of airflow vortex
shedding, the drop appears as quasi-steady. In fact, measurements by
Simon \cite{Simon-MS-21} show this exactly. Simon replaced a water-drop
with an equal-volume solid hemisphere and observed equivalent airflow
spectra in both cases.

In conclusion, although both large-amplitude drop interface motion and
airflow vortex shedding precede drop depinning, these phenomena appear
to be uncoupled. Thus, while modeling the unsteady dynamics of a
three-dimensional drop near critical conditions will remain a challenging task, it
appears as if a successful model will not need to include airflow
unsteadiness. On the time scale of key drop oscillation frequencies, the
airflow appears quasi-steady.